\documentclass[11pt,oneside,letterpaper]{article}
\pdfoutput=1

\usepackage[lmargin=70pt,rmargin=70pt]{geometry}

\usepackage{setspace}
\usepackage{graphicx}
\usepackage{subcaption}
\usepackage{amsmath}
\usepackage{amssymb}
\usepackage{physics}
\usepackage{color}
\definecolor{darkgreen}{rgb}{0,0.5,0}
\definecolor{darkblue}{rgb}{0,0,0.6}
\definecolor{purple}{rgb}{0.4,.2,0.7}
\usepackage[colorlinks=true,citecolor=darkgreen,linkcolor=darkblue,urlcolor=purple]{hyperref}
\usepackage{comment}
\usepackage{cite}

\newcommand{\dho}{\partial}

\newcommand{\cO}{\ensuremath{\mathcal{O}}}

\newcommand{\cD}{\ensuremath{\mathcal{D}}}

\renewcommand{\t}{\tau}

\DeclareMathOperator{\sign}{sign}

\begin{document}

\begin{center}
 \ \\
\vspace{1in}

{ \Large {\bf Does the SYK model have a spin glass phase?}}

\vspace{1cm}

Guy Gur-Ari$^{a,c}$, Raghu Mahajan$^{a,b}$ and Abolhassan Vaezi$^{c}$

\vspace{1cm}

{\small{\it 
$^a$ School of Natural Sciences, Institute for Advanced Study, \\
Princeton, New Jersey, USA\\
$^b$ Department of Physics, Princeton University, \\
Princeton, New Jersey, USA \\
$^c$ Department of Physics, Stanford University, \\
Stanford, California, USA \\
\vspace{0.3cm}
 }}

\vspace{1.6cm}

\end{center}

\begin{abstract}
  We argue that the Sachdev-Ye-Kitaev model has no spin glass phase, based on calculations involving both the nearly-conformal limit and the strongly-coupled Schwarzian limit of the model.
  This conclusion is supported by numerical computations of eigenvalue statistics with up to 46 Majorana fermions. 
  In addition, we find numerically that the distribution of the ground state energy is Gaussian.
\end{abstract}

\pagebreak
\setcounter{page}{1}

\tableofcontents{}

\section{Introduction and summary}

The Sachdev-Ye-Kitaev (SYK) model is a disordered quantum mechanical model of $N$ Majorana fermions that has remarkable properties, involving both its direct quantum mechanical description and its holographic dual \cite{Sachdev:1992fk,PhysRevB.59.5341, kitaevtalks, Maldacena:2016hyu, Kitaev:2017awl, Maldacena:2016upp}.
In the large $N$, strongly coupled limit the model becomes solvable, yet it remains chaotic.
It has a master field reformulation that is evocative of a simple bulk description.
While our understanding of its holographic dual is incomplete, many of the model's low temperature properties are reproduced by Jackiw-Teiltelboim gravity \cite{Kitaev:2017awl, Maldacena:2016upp, Almheiri:2014cka, Engelsoy:2016xyb, Mandal:2017thl, Jensen:2016pah}.
In particular, the infinite $N$ model has non-zero entropy at zero temperature and a maximal Lyapunov exponent \cite{kitaevtalks, Maldacena:2016hyu, Maldacena:2015waa}, two properties that are consistent with a bulk description involving an extremal black hole.
The SYK model has also found condensed matter applications in strongly-coupled transport and  entanglement dynamics \cite{Gu:2016oyy, Gu:2017ohj, Huang:2017nox, Gu:2017njx, Chowdhury:2018sho, Eberlein:2017wah, Davison:2016ngz}.
Finally, the fact that the model has a finite-dimensional Hilbert space at finite $N$ allows for straightforward and precise numerical computations.

It is natural to ask whether the model has a transition to a spin glass phase at low temperature---a common occurrence in disordered systems.
Indeed, the original Sachdev-Ye (SY) model \cite{Sachdev:1992fk} shares many properties with the SYK model, but in some versions of that model a spin glass transition occurs at a relatively high temperature \cite{PhysRevB.63.134406}.
A spin glass transition in the SYK model would imply a breakdown of the dual black hole picture.\footnote{
  The potential role of spin glass physics in quantum gravity was discussed in \cite{Denef:2011ee, Anninos:2011kh, Anninos:2012gk}.
  The notion of AdS$_2$ fragmentation \cite{Maldacena:1998uz} may also be relevant for holographic duals of quantum mechanical systems in a spin glass phase.
  }

We study this question by looking for two distinct signatures of a spin glass phase.
In both cases we find no indication of a spin glass transition, suggesting that the SYK model remains in its well-known paramagnetic phase down to arbitrarily low temperatures.

The first diagnostic is the condensation of replica off-diagonal modes.
In the ordinary high-temperature phase, the path integral is dominated by a saddle point that is both diagonal and symmetric in replica space.
A deviation from this would indicate a spin glass phase transition.
We compute the effective potential for some replica off-diagonal modes in the nearly-conformal limit of the theory, $1 \ll \beta J \ll N$ (here $\beta$ is the inverse temperature and $\beta J$ is the effective coupling; see Appendix~\ref{app:model} for our conventions).
The authors of \cite{PhysRevB.63.134406} carried out a similar calculation in the SY model, and for their `slave fermion' model they found a critical temperature $T_c \simeq J e^{-c\sqrt{N}}$, where $c$ is an order one constant.
For the SYK model, a similar comment was made in \cite{Kitaev:2017awl}.
We reproduce this estimate of $T_c$ using the conformal limit of the SYK model and extend it to arbitrary values of $q$, the order of the fermion interaction.

Such an exponentially low temperature lies outside the regime of validity of the conformal calculation.
Instead, the critical temperature falls within the strongly-coupled Schwarzian limit of the theory, namely $1 \ll N \ll \beta J$.
We repeat the calculation in the Schwarzian theory and find that the effect disappears: The off-diagonal modes we consider are always stable, indicating that there is no spin glass transition.
Our analytic results are presented for general $q$. However, as explained in the text, they are non-trivial only for $q\equiv_4 0$.
The reason is that for the other $q$ values the off-diagonal operator we are considering can never condense.

As a second diagnostic for a spin glass transition, we look for a deviation from Random Matrix Theory (RMT) predictions for the level-spacing statistics at low energies \cite{Guhr:1997jw}.
When the system is in a spin glass phase it loses ergodicity.
As a result, we expect its accessible energy states to become uncorrelated, and the level-spacing statistics to no longer follow RMT predictions.
In this work we present numerical results for the SYK model with up to $N=46$ Majorana fermions and with $q=4$.
These results were obtained by computing the lowest lying eigenvalues of the Hamiltonian on a cluster of GPUs.
Our results are all consistent with RMT predictions, and rule out a spin glass phase for all values of $N$ we tested.

The paper is organized as follows.
In Section~\ref{sec:an} we carry out the calculation involving the replica off-diagonal modes.
In Section~\ref{numerics} we present numerical results for the SYK model, testing RMT predictions involving level-spacing statistics, as well as the distribution of the ground state energy.
Several appendices expand on key points.
Appendix~\ref{app:model} includes our conventions and a brief review of the SYK model.
Appendix~\ref{app:deriveGqby2} includes details of the analytic calculation, and Appendix~\ref{app:numerics} describes the numerical methods used in this work.
Finally, Appendix~\ref{app:sk} reviews the relation between level-spacing statistics and a spin glass phase in the quantum Sherrington-Kirkpatrick model.

\section{Analytic results}
\label{sec:an}

In this section we present an analytic argument against a low-temperature spin glass phase in the SYK model.
See Appendix~\ref{app:model} for a brief review of the model.

In the $n$-replica theory, the condensation of a replica off-diagonal mode signals a spin glass transition.
Such condensation happens when the effective potential of the mode becomes unstable. 
The effective potential can be computed in the high temperature phase, which is the usual paramagnetic phase described by a replica-diagonal and replica-symmetric saddle point.

In the nearly-conformal limit, we find a predicted spin glass transition (for $q=4$) at a temperature $T_c \simeq J e^{-c\sqrt{N}}$ with some $c>0$.
Similar calculations were performed in \cite{0022-3719-13-24-005} for the quantum Sherrington-Kirkpatrick (SK) model, and in \cite{PhysRevB.63.134406} for the Sachdev-Ye (SY) model.
The predicted transition occurs at a temperature $\beta J \gg N$, which is outside the regime of validity of the conformal approximation. 
Instead, this temperature falls within the strongly-coupled regime of the Schwarzian theory.
We repeat the calculation in the Schwarzian theory, and find that the instability actually does not occur.

While these results provide evidence that a spin glass transition does not occur, they do not prove it conclusively.
For example, the presence of diagonal, replica-symmetry-breaking solutions may also signal such a transition, and we do not rule out such solutions analytically.

\subsection{Replica off-diagonal modes}

We now introduce the replica off-diagonal modes that will be the focus of the rest of this section, and write down their effective potential to second order in the fields.
Let us introduce $n$ replicas, labeled by $a,b=1,\dots,n$, and write down the partition function of the replicated theory.
After taking the disorder average, we find
\begin{align}
  \langle Z^n \rangle = \int \! \cD\psi \, 
  \exp \Bigg[
    &- \frac{1}{2} \int \! d\tau \, \psi_i^a \dho_\tau \psi_i^a
     + \frac{J^2}{2 q N^{q-1}}
    \sum_{a,b=1}^n \sum_{i_1,\dots,i_q} 
    \int \! d\tau_1 d\tau_2 \,
    \psi_{i_1}^a(\tau_1) \psi_{i_1}^b(\tau_2) \cdots \psi_{i_q}^a(\tau_1) \psi_{i_q}^b(\tau_2) 
  \Bigg] \,. \label{Zn}
\end{align}
Let us introduce the Hubbard-Stratonovich field $F_{ab}(\tau_1,\tau_2)$.
\begin{align}
  \langle Z^n \rangle = \int \! \cD\psi \, \cD F
  \exp \Bigg[
    &- \frac{1}{2} \int \! d\tau \, \psi_i^a \dho_\tau \psi_i^a
    - \frac{qN}{2J^2} \int d\tau_1 d\tau_2 F_{ab}^2(\tau_1,\tau_2)
    \cr &+
    N \int \! d\tau_1 d\tau_2 \, F_{ab}(\tau_1,\tau_2)
    \left( \frac{1}{N} \sum_i \psi_{i}^a(\tau_1) \psi_{i}^b(\tau_2) \right)^{q/2}
  \Bigg] \,. 
  \label{ZnF}
\end{align}
Note that this presentation of the theory in terms of the field $F$ is different than the common presentation in terms of Hubbard-Stratonovich fields $G$ and $\Sigma$ \cite{Maldacena:2016hyu}.
The saddle point equation for $F$ is
\begin{align}
  F_{ab}(\tau_1,\tau_2) &= \frac{J^2}{q} \left<
    \left( \frac{1}{N} \sum_i \psi_i^a(\tau_1) \psi_i^b(\tau_2) \right)^{q/2}
    \right> \,. \label{Fsaddle}
\end{align}
The correlator on the right-hand side is computed in the ordinary SYK theory.
We start in the usual (high temperature) phase, dominated by the known replica-diagonal saddle point.
In order to detect the putative spin glass phase we will lower the temperature, and look for an instability in modes $F_{ab}(\tau_1,\tau_2)$ with $a \ne b$.
If any of these modes condense, that is a signal of a spin glass transition.

For the $q=4$ theory, $F_{ab}$ is a 4-fermion operator.
Taking $a \ne b$, this is the minimal replica off-diagonal operator which can condense.
To see why, imagine computing $\langle \psi_a \psi_b \rangle$ where $\psi_a \psi_b$ is (schematically) a replica off-diagonal operator.
Suppose we do this by first computing the fermion path integral, followed by the disorder average.
In the first step the replicas are decoupled, and the calculation factorizes as $\langle \langle \psi_a \rangle_\psi \langle \psi_b \rangle_\psi \rangle_J$ into fermion 1-point functions, which vanish. Here $\langle \cdot \rangle_\psi$ denotes a fermion path integral and $\langle \cdot \rangle_J$ denotes disorder averaging.

For the theories with $q=2,6,10,\dots$, the off-digonal operators $F_{ab}$ involve an odd number of fermions in each replica.
The same argument then shows that these operators cannot condense, and we expect their effective potentials to always be stable.
For theories with $q=8,12,\dots$ an instability in $F_{ab}$ is possible, but our $F_{ab}$ is not the minimal operator that can condense as it involves more than 4 fermions.
Therefore, while we carry out the calculation for general $q$, the resulting evidence against a spin glass transition only applies to theories with $q \equiv_4 0$ and is strongest for $q=4$.

We focus on the time-independent modes $F_{ab}$ with $a \ne b$.
In Appendix~\ref{app:deriveGqby2} we compute the quadratic piece in the effective potential of these modes, $V_{\rm eff}(F_{ab}) = \frac{1}{2} \beta^4 m_{ab}^2 F_{ab}^2 + \cO(F_{ab}^3)$, and find the following squared-mass.
\begin{align}
  \beta^4 m_{ab}^2 = \frac{qN}{J^2} \beta^2 -
  (-1)^{q/2} \left( q/2 \right)! \, N^{2-q/2}
  \left( \int_0^\beta \! d\t_{1} d\t_2 \, \Big\langle G^{q/2}(\t_1,\t_2) \Big\rangle \right)^2 \,.
  \label{mab}
\end{align}
Here $G(\tau_1, \tau_2)$ is the usual fermion bilinear operator, defined in Appendix \ref{app:model}.
The first term on the right-hand side is leading at large $N$ and fixed temperature.
A phase transition will happen if $m_{ab}^2$ becomes negative at sufficiently low temperature.
Again, the correlator appearing in \eqref{mab} is a correlator in the ordinary SYK theory.

Notice that an instability is only possible when $q \equiv_4 0$, consistent with the argument above. 
We will assume this from now on.
Let us now compute the effective mass in two different limits of the theory.

\subsection{Nearly-conformal limit}

Let us compute the squared mass \eqref{mab} in the nearly-conformal limit.
At leading order in large $N$ the correlator factorizes as $\langle G^{q/2} \rangle = \langle G \rangle^{q/2} + \cdots$.
In this limit, the correlator is given by
\begin{align}
  \langle G(\t) \rangle &=
  b \left( \frac{\pi/\beta}{\sin (\pi \tau/\beta)} \right)^{2/q} \sign(\tau) \,, \quad
  b^q =
  \frac{1}{\pi J^2} \left( \frac{1}{2} - \frac{1}{q} \right) \tan \left( \frac{\pi}{q} \right) \,.
\end{align}
Using this result, we find a log divergence in the integral that we regularize by introducing a cutoff $\epsilon$ on Euclidean time.
\begin{align}
  \int \! d\t_{1}  d\t_2 \, \langle G^{q/2}(\t_1,\t_2) \rangle
  &\simeq 
  \beta \int_\epsilon^{\beta-\epsilon} \! d\t  \, \langle G(\t) \rangle^{q/2}
  = - 2\, b^{q/2} \beta \log \tan \left( \frac{\pi \epsilon}{2\beta} \right) \,. \label{int}
\end{align}
We place the cutoff at $\epsilon \approx J$, where we expect the correlator to become free, $\langle G(\tau) \rangle_{\rm free} = \frac{1}{2} \sign(\tau)$.
With these approximations, we find
\begin{align}
  m_{ab}^2 &= \frac{1}{(\beta J)^2} \left(
    qN - a N^{2-q/2} \log^2 (\beta J)
  \right) \,,\quad
  a = \frac{4}{\pi} (q/2)! \left( \frac{1}{2} - \frac{1}{q} \right) \tan(\pi/q) \,. \label{mab2}
\end{align}
The squared-mass becomes negative at the critical temperature
\begin{align}
  T_c =
  J \exp \left[ - \left( \frac{q}{a} \right)^{1/2} N^{\frac{q-2}{4}} \right] \,,\quad
  q=4,8,12,\dots \,.
  \label{eq:tcconf}
\end{align}
For $q>4$, this predicted critical temperature is parametrically smaller than $e^{-N}$, the typical level spacing. 
Only the ground state is accessible at this temperature and so this result suggests there is no spin glass phase transition for these theories. 
For $q=4$, we get the predicted transition temperature
  \begin{align}
  T_c = J\, e^{- \sqrt{2\pi N} } \quad \text{for } q=4.
  \end{align}

Notice that $J/T_c \gg N$, and therefore this critical temperature lies outside the regime of validity of the conformal approximation.
The correct description at this temperature is the strongly-coupled Schwarzian theory.
We now turn to computing the critical temperature in this limit.

\subsection{Strongly-coupled Schwarzian limit}

The Schwarzian theory is a solvable theory with an inverse coupling constant $C = N \alpha_S(q)/J$ \cite{Stanford:2017thb, Bagrets:2016cdf, Bagrets:2017pwq, Mertens:2017mtv, Mertens:2018fds, Maldacena:2016hyu}; see Appendix~\ref{app:model} for a brief review.
In this section we assume that we are in the strong coupling limit $C \ll \beta$.
We use the results of \cite{Mertens:2017mtv,Yang:2018} to compute the correlator in the Schwarzian theory.\footnote{
  See for example equation (4.10) in \cite{Mertens:2017mtv}.
  In their notation, we set $\ell=1/2$ (the dimension of the operator $G^{q/2}$).
  We thank Zhenbin Yang for sharing an early draft of \cite{Yang:2018}.
}
\begin{align}\
    \langle G^{q/2}(\tau) \rangle = \frac{c_1}{N}\frac{(\beta/C)^{3/2}}{e^{2\pi^2 C/\beta}} 
    \int dk_1^2 dk_2^2\,
    \frac{\sinh(2\pi k_1) \sinh(2\pi k_2)}{\cosh(2\pi k_1) + \cosh(2\pi k_2)} 
    \exp\!\left( -\frac{\tau k_1^2}{2C} \right) 
    \exp\!\left( -\frac{(\beta - \tau) k_2^2}{2C} \right) \,. \label{schcorr}
\end{align}
Here $c_1$ is a constant whose precise value will not be important to us.
The factor $(\beta/C)^{3/2}\exp(-2\pi^2 C/\beta)$ comes from the normalization by $1/Z$ \cite{Yang:2018,Stanford:2017thb}.
We will analyze this formula in two limits.

Let us first consider the regime $ \tau \ll C$. 
The factor $\exp(-(\beta-\tau)k_2^2/2C)$ is only significant for 
$k_2 \lesssim \sqrt{C/\beta} \ll 1$. 
This means that to leading order we can set $k_2 = 0$ in the $\cosh(2\pi k_2)$ in the denominator.
The $k_1$ integral is dominated by the range $k_1 \gtrsim 1$, where we can approximate $\sinh(2\pi k_1)/(1+\cosh(2\pi k_1)) \approx 1$.
Ignoring overall numerical coefficients, we get
\begin{align}
  \langle G^{q/2}(\tau)  \rangle &\simeq \frac{(\beta/C)^{3/2}}{N}
  \int_0^\infty dk_2^2\, k_2 \exp(- \frac{\beta k_2^2}{2C}) 
  \int_0^\infty dk_1^2\,  \exp(-\frac{\tau k_1^2}{2C})  \\
  &\simeq \frac{1}{J \tau} \qquad \text{for } \tau \ll C.
  \label{Gq2reg2}
\end{align}
Note that we recovered the conformal answer.

Next, consider the regime $ C \ll \tau \leq \beta/2$. 
Now both $\tau,\beta-\tau \gg C$ and the integral in \eqref{schcorr} is dominated by the region $k_1 \lesssim \sqrt{C/\tau} \ll 1$ and $k_2 \lesssim \sqrt{C/(\beta - \tau)} \ll 1$.
Thus we have
\begin{align}
  \langle G^{q/2}(\tau) \rangle &\approx \frac{(\beta/C)^{3/2}}{N}\int_0^\infty dk_1^2 k_1 \exp \left( \frac{-\tau k_1^2}{2C} \right)
  \int_0^\infty dk_2^2 k_2 \exp\left( \frac{-(\beta - \tau) k_2^2}{2C}\right) \nonumber \\
  &\simeq \frac{1}{N}
  \left[ \frac{\beta C}{\tau(\beta-\tau)} \right]^{3/2} 
  \qquad \text{for } \tau \gg C.
  \label{Gq2reg3}
\end{align}

Let us analyze how the effective mass (\ref{mab}) changes compared to the conformal answer (\ref{mab2}).
The negative contribution to the effective mass (\ref{mab}) is proportional to $\left( 2\beta \int_0^{\beta/2} d\tau \langle G^{q/2}(\tau) \rangle \right)^2$. 
Here we used the fact that $\langle G(\tau) \rangle$ is symmetric about $\tau = \beta/2$. 
Let us compute the $\tau$ integral by splitting it into three regions: $\tau \in (0, 1/J)$, $(1/J, C)$ and $(C, \beta/2)$.
In the first region, the correlator is approximately equal to the free correlator which is a constant, and we get a contribution proportional to $1/J$.
From the second region we get, using \eqref{Gq2reg2},
\begin{align}
    \int_{1/J}^{C} \frac{d\tau}{J\tau} \simeq \frac{\log CJ}{J} \simeq \frac{\log N}{J}\, .
    \label{Gintreg2}
\end{align}
The contribution of the third region is, using \eqref{Gq2reg3},
\begin{align}
    \frac{(\beta C)^{3/2}}{N}\int_C^{\beta/2} \frac{d\tau}{\tau^{3/2}(\beta - \tau)^{3/2}} 
 \simeq \frac{C}{N} \simeq \frac{1}{J}\, .
\end{align}
We see that the dominant contribution to the $\tau$ integral at large $N$ is from the second region, equation \eqref{Gintreg2}.
Thus from (\ref{mab}), now setting $q=4$ for simplicity, we have
\begin{align}
    m_{ab}^2 = \frac{4N - c_2 \log^2 \!N}{(\beta J)^2} \,.
\end{align}
Here $c_2$ is a positive constant.
Comparing to the conformal answer (\ref{mab2}), we see that $\log (\beta J)$ got replaced by $\log N$.
Thus the effective mass is always positive and the mode is stable.
The same conclusion holds for other values of $q$.

\section{Numerical results}
\label{numerics}

In the previous section we studied the condensation of replica off-diagonal modes, which serve as a signature for a spin glass phase transition.
For quantum systems, the level spacing statistics are another such signature.
Indeed, in an ordinary chaotic system the level spacing statistics obey Random Matrix Theory (RMT) predictions, implying for example level repulsion \cite{Guhr:1997jw}; in a spin glass the levels are decorrelated and there is no level repulsion.
These relations are reviewed in Appendix~\ref{app:sk} for the quantum Sherrington-Kirkpatrick model.

In this section we present numerical results for the spectrum and level spacing statistics of the SYK model with 4-fermion interactions.
These results were computed by partially diagonalizing the Hamiltonian, obtaining the energy levels at the edge of the spectrum.
Details about the numerical methods used here can be found in Appendix~\ref{app:numerics}.
Our level spacing results exhibit RMT behavior down to the lowest observed energies, and these results favor our conclusion that the model has no spin glass phase transition at low temperature.
In addition, we find numerically that the ground state energy follows a Gaussian distribution.

\subsection{The edge of the spectrum}
\label{sec:edge}

\begin{figure}[t]
  \centering
  \begin{subfigure}[t]{0.48\textwidth}
    \centering
    \includegraphics[width=0.9\textwidth]{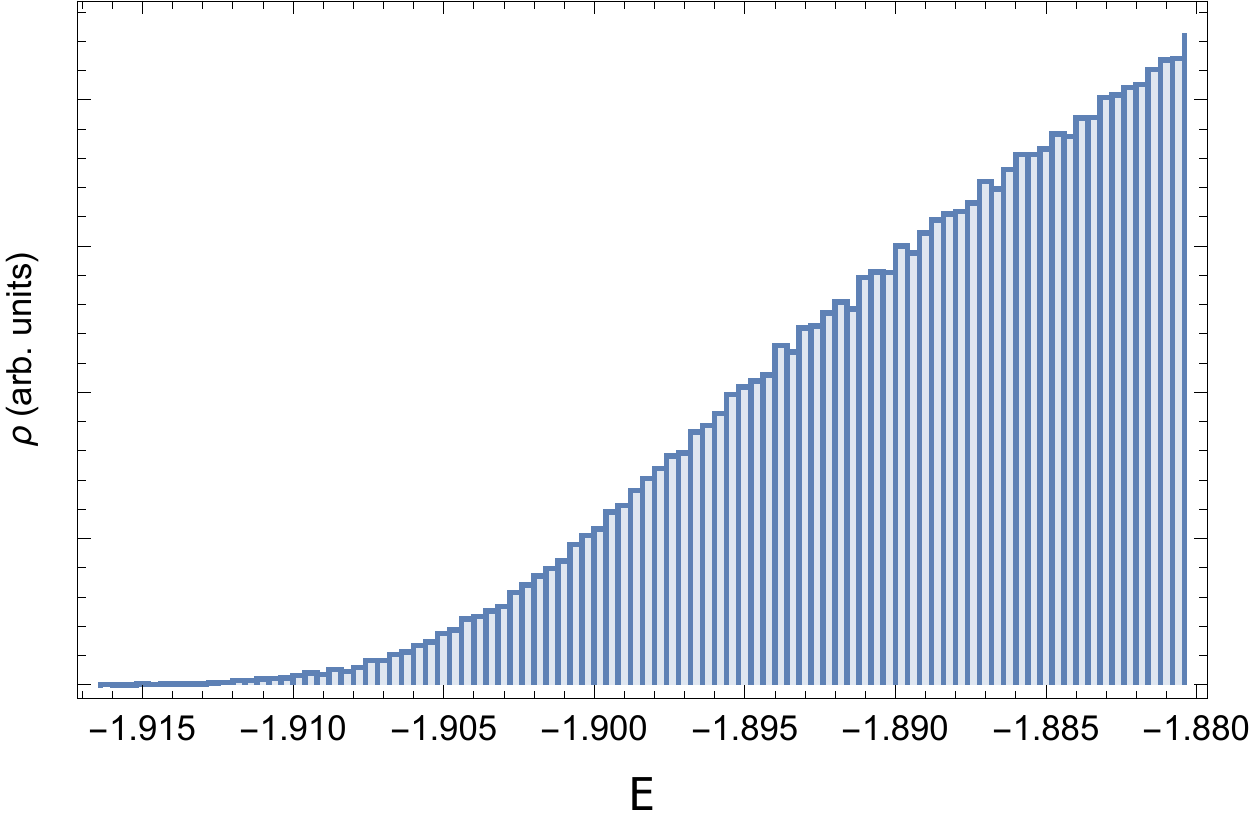}
    \caption{}
  \end{subfigure}
  ~ 
  \begin{subfigure}[t]{0.48\textwidth}
    \centering
    \includegraphics[width=0.9\textwidth]{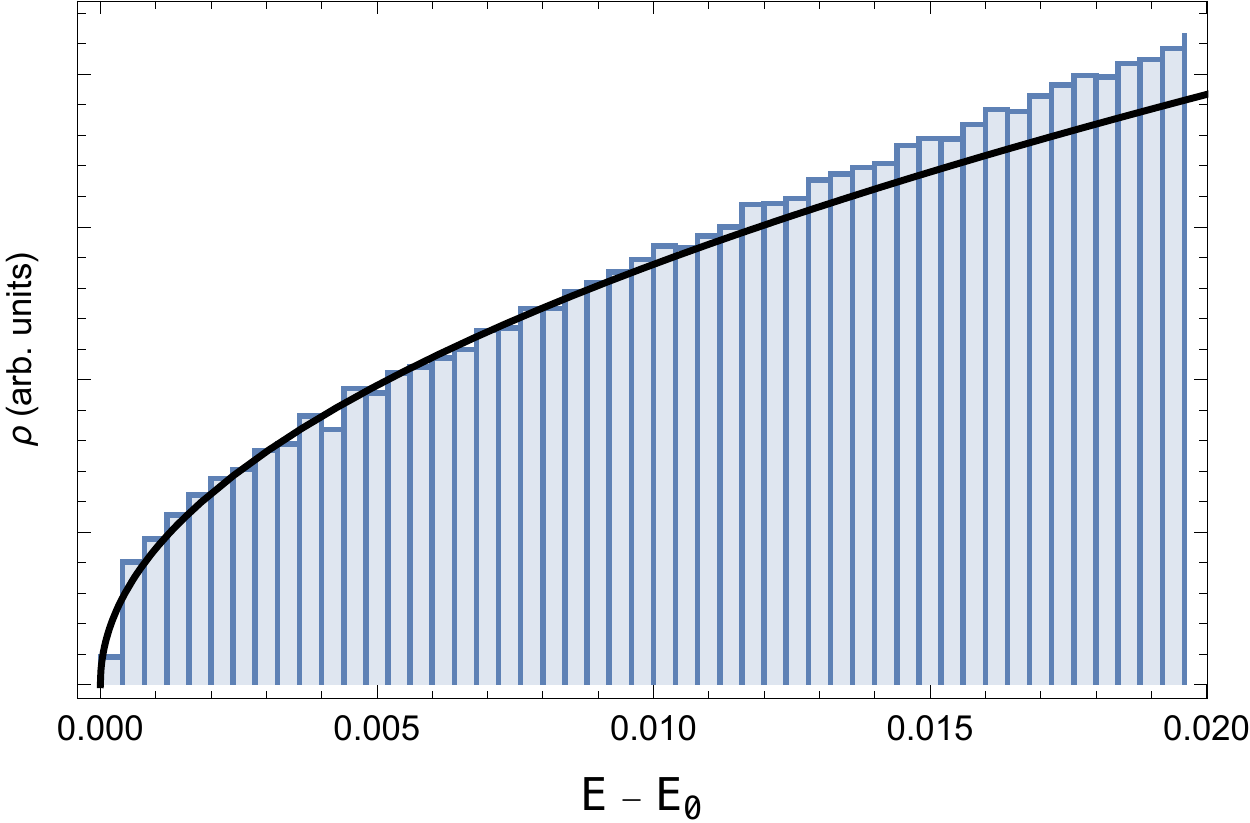}
    \caption{}
  \end{subfigure}
  \caption{The density of states for SYK near the edge of the spectrum, with $N=42$ Majorana fermions and 800 realizations, each with about 1,200 eigenvalues at each edge of the spectrum.
    (a) The density of states. (b) The density of states, with the energies of each realization shifted by its respective ground state energy.
    The fit is to a power law for the range $E-E_0<0.01$, and gives $\rho \sim (E-E_0)^{0.49}$.
    (The best-fit exponent varies between $0.4-0.6$ depending on the choice of range.)
  }
  \label{fig:spec}
\end{figure}

Figure~\ref{fig:spec} shows the spectral density at the edge of the spectrum.\footnote{
  In this work we treat the high edge of the spectrum as an independent realization (it corresponds to the low edge of the spectrum for the realization with all random couplings negated).
  This doubles our effective number of realizations, and we quote this effective number.
}
At large $N$ and low energies, the analytic prediction \cite{Maldacena:2016hyu, Stanford:2017thb} is that the density should behave as $\rho(E) \sim \sqrt{E-E_0}$ near the edge.
If we simply plot the energy density, we find that there are large fluctuations that mask this effect.
However, if we shift the energies of each realization by its respective ground state energy (such that the ground state energy of each realization becomes zero), the predicted edge behavior becomes clearly visible.

\begin{figure}[ht!]
  \centering
  \includegraphics[width=0.5\textwidth]{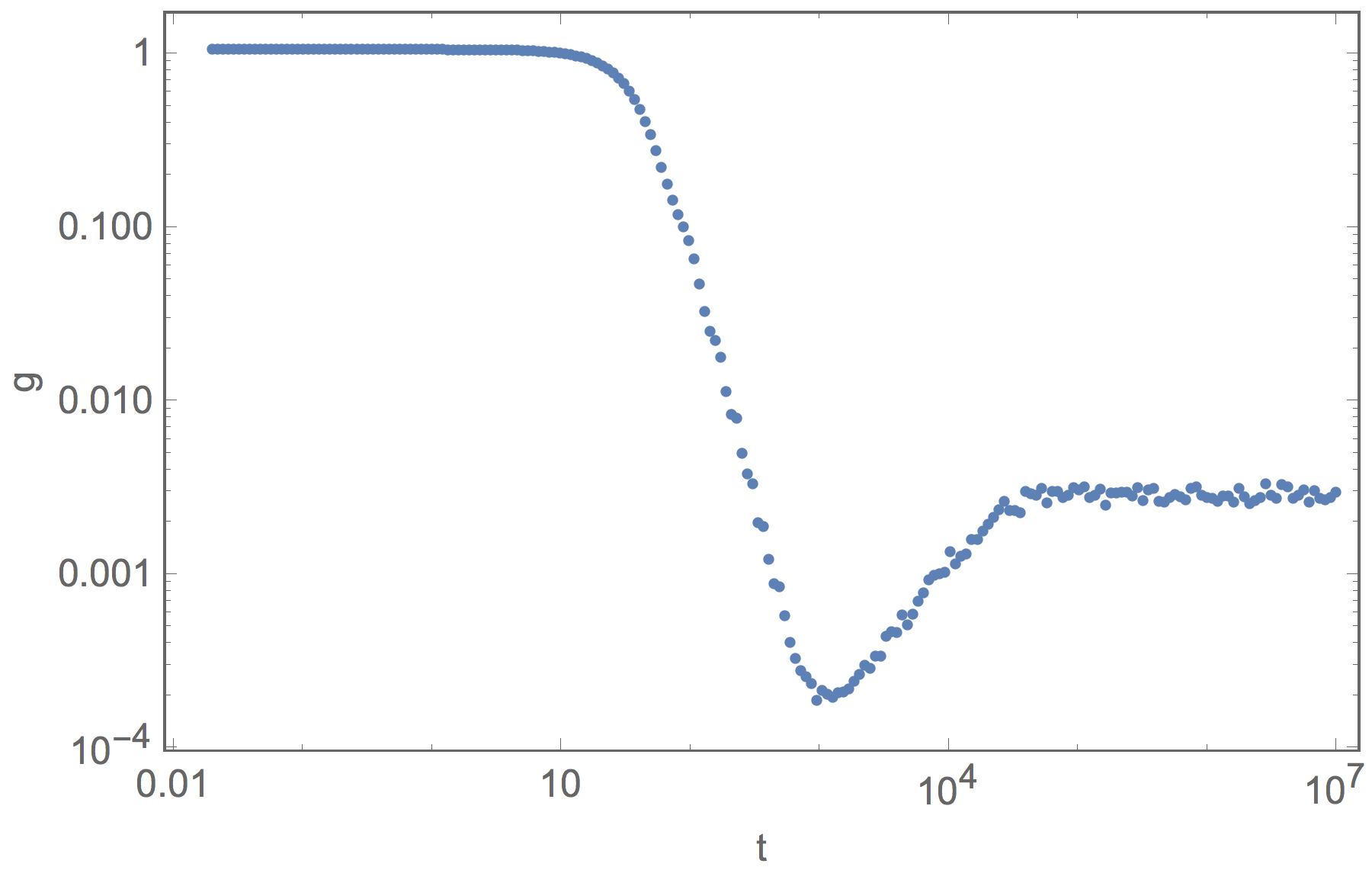}
  \caption{The Spectral Form Factor (denoted $g$ in \cite{Cotler:2016fpe}) with $N=42$ Majorana fermions and $\beta=50$, using the same data as in Figure~\ref{fig:spec}.}
  \label{fig:sff}
\end{figure}

In \cite{Cotler:2016fpe}, the spectral form factor was introduced as a diagnostic of the late-time dynamics, with connections both to the information paradox and to RMT; see also the recent paper \cite{Saad:2018bqo}.
Our numerical results allow us to test the RMT predictions at larger values of $N$ and at lower temperatures.
Figure~\ref{fig:sff} shows these results.
The three notable features discussed in \cite{Cotler:2016fpe}, the early `slope' followed by the late time `ramp' and `plateau', are clearly visible.
In particular, the ramp is consistent with RMT predictions and indicates a chaotic spectrum.

\subsection{Level spacing statistics}
\label{sec:lss}

In order to determine the phase of the system at low energies, we compute the level spacing statistics near the edge of the spectrum.
Statistics that agree with Random Matrix Theory predictions imply a chaotic phase, while statistics that follow an exponential distribution (corresponding to uncorrelated energy levels) are a signature of a spin glass phase. See Appendix~\ref{app:sk} for further discussion.

The standard method of computing level spacing statistics involves first `unfolding' the energy levels such that the mean energy density is one (see for example \cite{Guhr:1997jw}).
This procedure works well in the bulk of the spectrum, but becomes unreliable near the edge due to large fluctuations (such as the ones described in Section~\ref{sec:edge}).
We compute the level spacing statistics in two different ways that sidestep this problem.
First, we compute the level spacing distribution for the two lowest energy levels, collecting statistics only over different realizations.
Second, we compute the distribution of $\log(r_n)$ where $r_n = (E_n - E_{n+1}) / (E_{n+1} - E_{n+2})$ for a fixed number of lowest energy states.
Both of these distributions can be compared directly with RMT predictions without unfolding \cite{You:2016ldz}.

\begin{figure}[ht!]
  \centering
  \begin{subfigure}[t]{0.48\textwidth}
    \centering
    \includegraphics[width=0.92\textwidth]{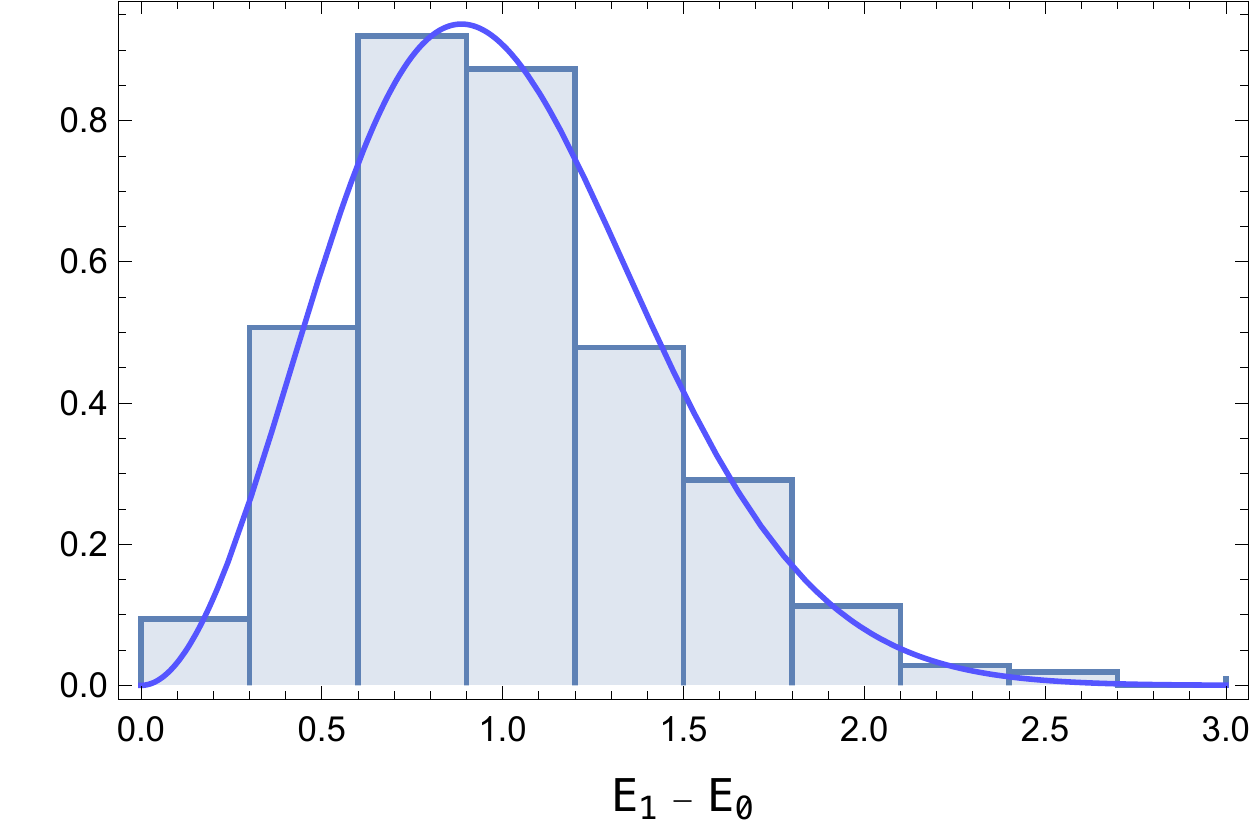}
    \caption{}
  \end{subfigure}
  ~ 
  \begin{subfigure}[t]{0.48\textwidth}
    \centering
    \includegraphics[width=0.9\textwidth]{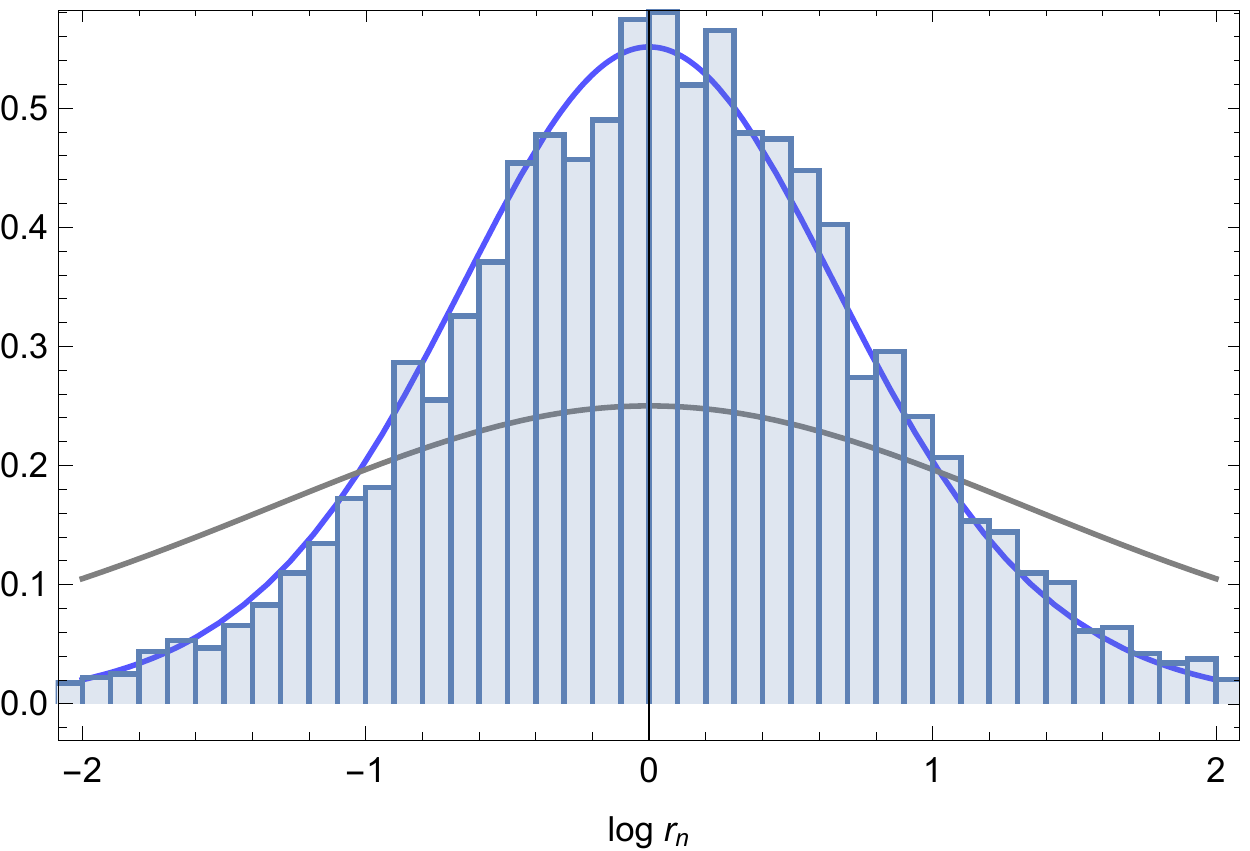}
    \caption{}
  \end{subfigure}
  \caption{Level spacing statistics for the edge of the SYK spectrum, with $N=46$ Majorana fermions and 355 realizations.
    (a) The spacing distribution for the two lowest levels, compared with the RMT prediction for the corresponding GUE ensemble.
    (b) The distribution of $\log(r_n)$ (described in the text) computed over the lowest 20 energy levels, compared with the RMT prediction (blue) and with the prediction for uncorrelated energies (gray).
  }
  \label{fig:ls}
\end{figure}

The results, shown in Figure~\ref{fig:ls}, are all consistent with RMT predictions.
We compare the computed level-spacing distribution against the Wigner surmise.
At the edge of the random matrix spectrum, the eigenvalue density correlations are described by the Airy kernel \cite{bowick1991universal,forrester1993spectrum,tracy1994level}, while in the bulk of the spectrum they are described by the sine kernel.
Despite this difference, it is easy to check empirically that the RMT nearest-neighbor level statistics are well approximated by the Wigner surmise in both cases.

Our results rule out a spin glass phase for the SYK model with up to $N=46$ Majorana fermions.
The ordinary paramagnetic phase persists down to arbitrarily low energies, even when the thermodynamic approximation breaks down and it is not useful to discuss temperatures.

\subsection{Ground state energy distribution}

In this section we compute the ground state energy distribution of the model (this was previously studied in \cite{Garcia-Garcia:2017pzl}).
The extremal eigenvalues of matrices in common Random Matrix Theory ensembles follow a Tracy-Widom distribution \cite{tracy2002distribution,tracy2009distributions}.
In light of the detailed agreements between SYK and RMT described above, it is natural to ask whether the ground state energy distribution is also consistent with RMT predictions.
We observe numerically that this is not the case, and instead the ground state energy follows a Gaussian distribution.
This result is not surprising, as the RMT predictions for extremal eigenvalues are known to apply less universally than the predictions related to level spacing statistics.

\begin{figure}[ht]
  \centering
  \includegraphics[width=0.7\textwidth]{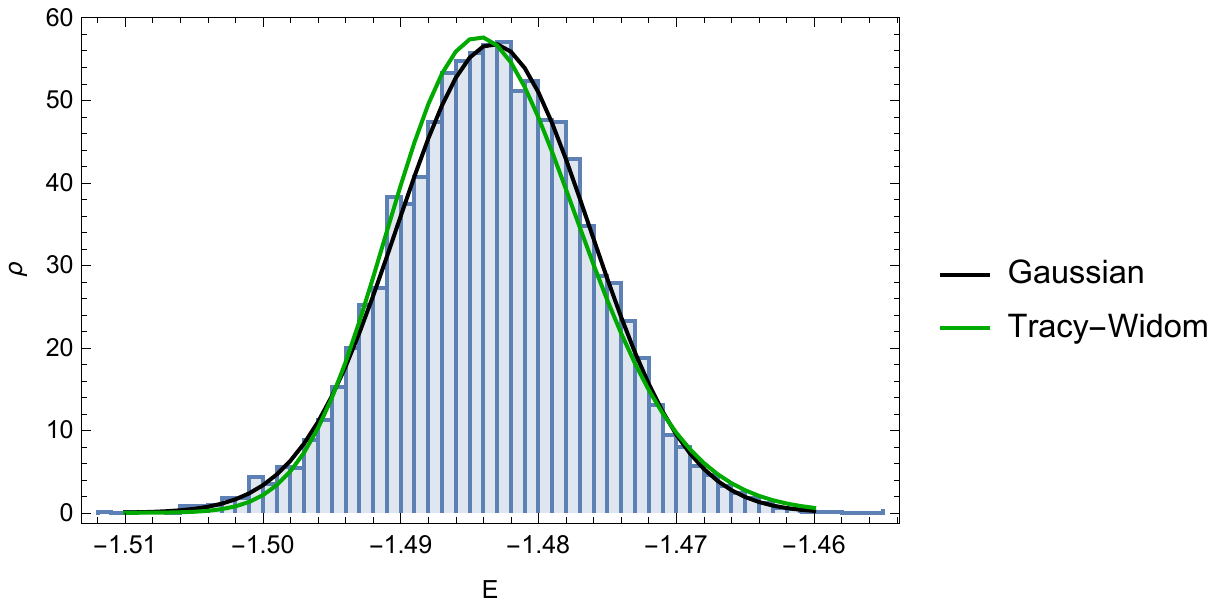}
  \caption{The ground state distribution for SYK with $N=32$ Majorana fermions, with statistics collected over $10^4$ realizations. Solid lines show the Gaussian and Tracy-Widom distributions with mean and variance chosen to fit the data.}
  \label{fig:gndHist}
\end{figure}

Figure~\ref{fig:gndHist} shows the ground state energy distribution, along with the Gaussian and Tracy-Widom distributions.
The mean and variance of both distributions were chosen to fit the data.
Just by eye, it is hard to determine which distribution fits the data better. 
We can distinguish the two distributions by considering higher order moments; in particular, the Tracy-Widom distribution is slightly skewed.\footnote{
  Recall that for a random variable $X$ with mean $\mu$ and variance $\sigma^2$, the skewness is defined by $\langle (X - \mu)^3/\sigma^3 \rangle$ and the kurtosis by $\langle (X - \mu)^4/\sigma^4 \rangle$.
  }
Table~\ref{tab:moments} lists these results, which show that the Gaussian distribution is clearly preferred.
\begin{table}[ht!]
  \centering
  \begin{tabular}{|l|cc|}
  \hline
    Distribution & Skewness & Kurtosis \\
    \hline
    Gaussian & 0 & 3 \\
    Tracy-Widom (GOE) & $-0.293$ & 3.165 \\
    Tracy-Widom (GUE) & $-0.224$ & 3.093 \\
    Tracy-Widom (GSE) & $-0.166$ & 3.049 \\
    \hline
    SYK $N=24$ (GOE) & $-0.070 \pm 0.020$ & $3.06 \pm 0.05$ \\
    SYK $N=26$ (GUE) & $-0.020 \pm 0.020$ & $3.02 \pm 0.05$ \\
    SYK $N=30$ (GUE) & $-0.021 \pm 0.019$ & $3.00 \pm 0.05$ \\
    SYK $N=32$ (GOE) & $-0.023 \pm 0.020$ & $3.03 \pm 0.05$ \\
    SYK $N=34$ (GUE) & $-0.007 \pm 0.019$ & $2.98 \pm 0.05$ \\
    \hline
  \end{tabular}
  \caption{Higher moments for the ground state energy distribution.
    SYK data was collected from $10^4$ realizations for each value of $N$.
  GSE data is not shown because the ground state is in the odd charge sector, and we only computed the even charge sector.}
  \label{tab:moments}
\end{table}

\begin{figure}[ht!]
  \centering
  \begin{subfigure}[t]{0.48\textwidth}
    \centering
    \includegraphics[width=0.87\textwidth]{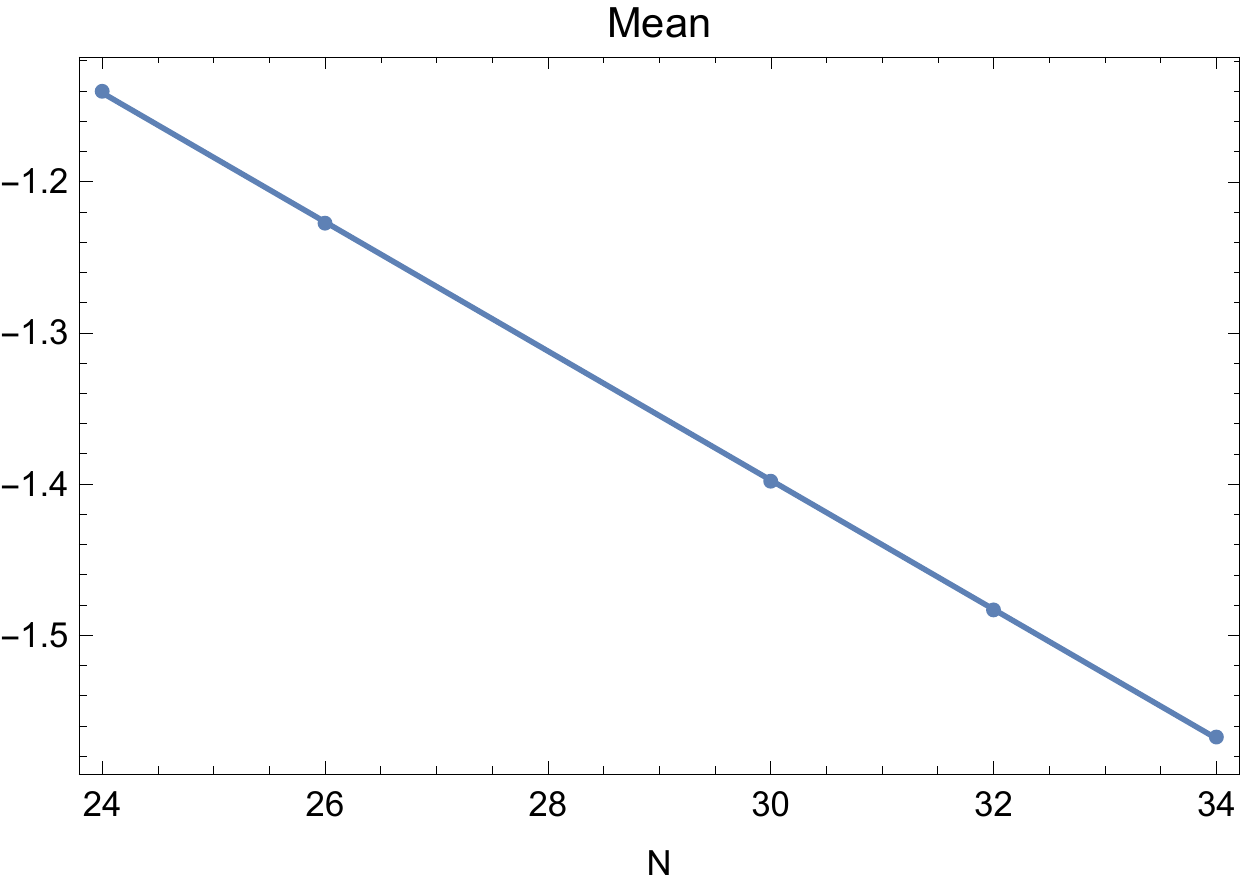}
    \caption{}
  \end{subfigure}
  ~ 
  \begin{subfigure}[t]{0.48\textwidth}
    \centering
    \includegraphics[width=0.9\textwidth]{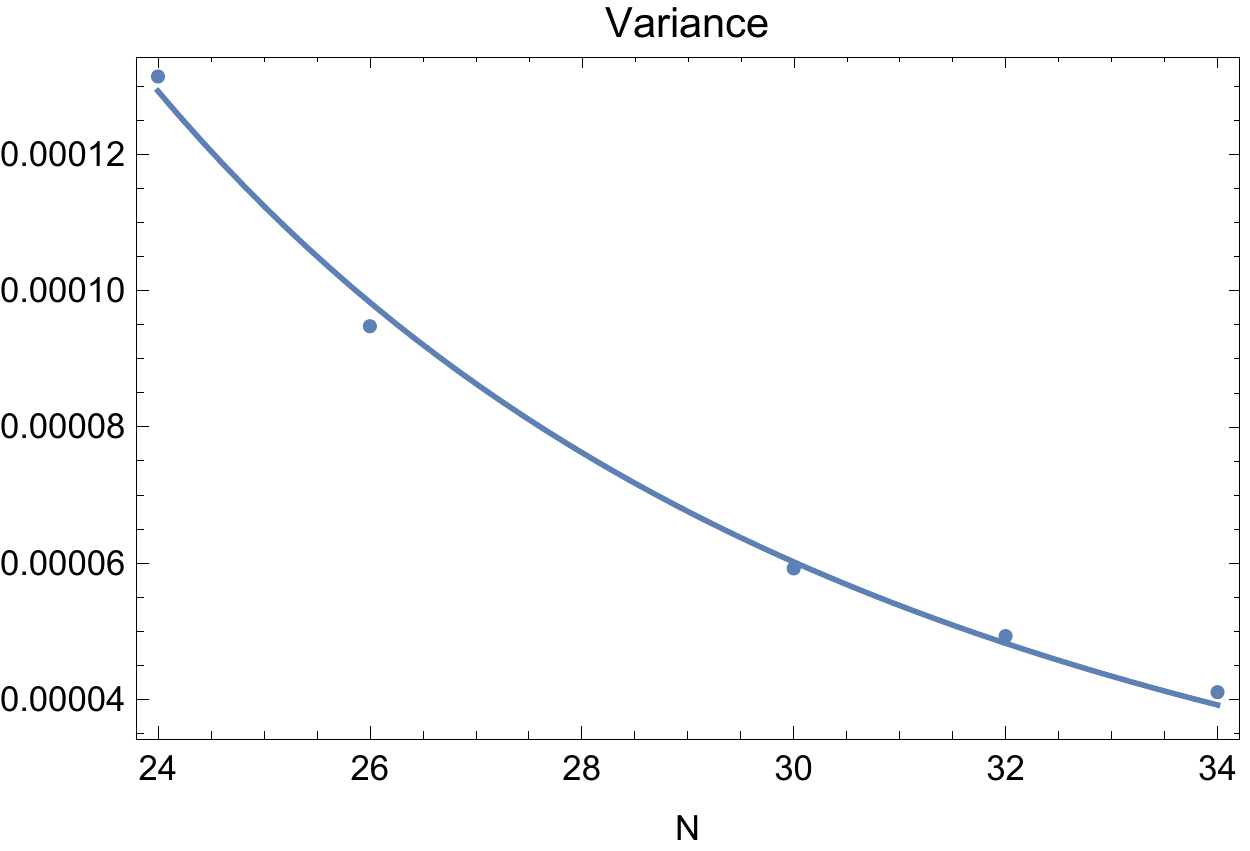}
    \caption{}
  \end{subfigure}
  \caption{Ground state mean and variance as a function of $N$, computed over $10^4$ realizations for each value of $N$.
    (a) The mean ground state energy, with a linear fit to $\langle E_0 \rangle = -0.043 N - 0.12$.
    (b) The variance of the ground state energy, with a power-law fit to $\mathrm{Var}(E_0) \sim N^{-3.43}$.
  }
  \label{fig:gndMV}
\end{figure}

Next, Figure~\ref{fig:gndMV} shows the dependence of the Gaussian parameters on $N$.
We find that the leading large $N$ term in the mean ground state energy is within 10\% of the analytic large $N$ prediction \cite{Maldacena:2016hyu}, which is $E_0 \approx -0.0406 N$.
For the variance we find that a power law provides a good fit.

\newpage
\section*{Acknowledgments}

The authors would like to Tarek Anous, David Huse, Alexei Kitaev, Juan Maldacena, Edgar Shaghoulian, Stephen Shenker, Douglas Stanford, Sho Yaida, and Zhenbin Yang for useful discussions.
G.G. was supported by a grant from the John Templeton Foundation, and is supported by NSF grant PHY-1606531.
The opinions expressed in this publication are those of the authors and do not necessarily reflect the views of the John Templeton Foundation.
R.M. is supported by US Department of Energy grant No. DE-SC0016244.
A.V. acknowledges the Gordon and Betty Moore Foundation’s EPiQS Initiative through Grant GBMF4302.

\appendix

\section{The SYK model}
\label{app:model}

In this section we briefly review some basic properties of the SYK model, following \cite{Maldacena:2016hyu}.
The dynamical degrees of freedom of the model are $N$ Majorana fermions $\psi_1, \ldots ,\psi_N$.
The Hamiltonian is
\begin{align}
  H = i^{q/2} \sum_{i_1<i_2<\cdots<i_q} J_{i_1 \cdots i_q} \psi_{i_1 \cdots i_q} \,.
\end{align}
Here $q$ is a positive even integer, and we introduced the notation $\psi_{i_1\cdots i_q} = \psi_{i_1} \psi_{i_2} \cdots \psi_{i_q}$. 
For each choice of $i_1<i_2<\cdots<i_q$, the coupling $J_{i_1 \ldots i_q}$ is an independent Gaussian random variable with zero mean and with variance given by
\begin{align}
  \langle J_{i_1 \cdots i_q}^2 \rangle = \frac{(q-1)! J^2}{N^{q-1}} \,. \label{varianceJ}
\end{align}
The Euclidean time action is
\begin{align}
  S = \int_0^\beta d\tau \left(
    \frac{1}{2} \psi_i \dho_\tau \psi_i
    - i^{q/2} \sum_{i_1<i_2<\cdots<i_q}
    J_{i_1 \cdots i_q} \psi_{i_1 \cdots i_q} \right) \,.
\end{align}
The fermion bilinear operator is defined as
\begin{align}
  G(\tau_1, \tau_2) = \frac{1}{N} \sum_{i=1}^N \psi_i(\tau_1) \psi_i (\tau_2)\, .
\end{align}

The nearly-conformal limit of the theory is $1 \ll \beta J \ll N$.
In this limit, the fermion 2-point function is given by 
\begin{align}
  \langle G(\t) \rangle &=
  b \left( \frac{\pi/\beta}{\sin (\pi \tau/\beta)} \right)^{2\Delta} \sign(\tau) \,, \label{GG}
  \\
   b^q &=
    \frac{1}{\pi J^2} \left( \frac{1}{2} - \Delta \right) \tan(\pi\Delta) \,,\quad \Delta =\frac{1}{q} \,. \label{bb}
\end{align}
This solution defines a replica-diagonal saddle point of the theory, written in terms of its master fields $G$ and $\Sigma$ \cite{Maldacena:2016hyu}.

The SYK action has an emergent time reparametrization symmetry $\tau \to f(\tau)$, which is spontaneously broken by the above solution (\ref{GG}).
Furthermore, this symmetry is explicitly broken by corrections to the conformal limit.  
The effective low-energy action of the theory, which governs the dynamics of the pseudo Nambu-Goldstone modes $f(\tau)$, is the Schwarzian action
\begin{align}
S = -C \int_0^\beta d\tau\, \text{Sch} \left( 
\tan \frac{f(\tau)}{2}, \tau
\right)\, ,
\end{align}
where $C = N \alpha_S(q) / J$.
Here $\alpha_S(q)$ is a numerical coefficient whose precise values can be found in \cite{Maldacena:2016hyu}.
In the weak coupling limit (corresponding to $\beta J \ll N$), the fluctuations about the saddle point $f(\tau) = \tau$ are small, and one can reproduce many results of the SYK model in the conformal limit.
In the strong coupling limit (corresponding to $\beta J \gg N$) the theory is still solvable \cite{Stanford:2017thb, Bagrets:2016cdf, Mertens:2017mtv}.

\section{Derivation of the effective mass}
\label{app:deriveGqby2}

In this appendix we derive equation \eqref{mab} for the effective mass of the replica off-diagonal modes $F_{ab}$.
The effective action for these modes is defined by
\begin{align}
  \int \cD F e^{-S_{\rm eff}(F)} = \langle Z^n \rangle \,,
  \label{Seffint}
\end{align}
where the replicated partition function was given in \eqref{ZnF}.
Expanding the effective action to quadratic order, we have
\begin{align}
  S_{\rm eff}(F) = 
  \frac{1}{2} \sum_{a,b,c,d} \int \! d\t_{1,2,3,4} \, m_{ab,cd}^2(\t_1,\t_2,\t_3,\t_4) F_{ab}(\t_1,\t_2) F_{cd}(\t_3,\t_4) +
  \cO(F_{ab}^3) \,.
\end{align}
We show below that only terms in which $(a,b)=(c,d)$ have non-trivial masses, and compute the effective squared-mass $m_{ab}^2 \equiv m^2_{ab,ab}$ of the time-independent modes.

Expanding equations \eqref{Seffint} to second order in $F_{ab}(\tau_1, \tau_2)$, and using \eqref{ZnF}, we get
\begin{align}
  - \frac{1}{2} \sum_{a,b,c,d} \int m_{ab,cd}^2 F_{ab} F_{cd} &=
    - \frac{qN}{2J^2} \sum_{a,b} \int d\t_1 d\t_2 F_{ab}^2(\t_1,\t_2)
    \cr &\quad
    + \frac{1}{2} \left\langle \left\{
    N \sum_{a,b} \int \! d\t_1 d\t_2 \, F_{ab}(\t_1,\t_2)
    \left( \frac{1}{N} \sum_i \psi_{i}^a(\t_1) \psi_{i}^b(\t_2) \right)^{q/2}
  \right\}^2 \right\rangle \,. \cr \label{Fquad}
\end{align}
The second term on the right-hand side can be written as
\begin{align}
    \frac{N^{2-q}}{2} \sum_{a,b,c,d} \int \! d\t_1 d\t_2 d\t_3 d\t_4 \,
    F_{ab}(\t_1,\t_2) F_{cd}(\t_3,\t_4)
    \left\langle
      \left( \sum_i \psi_{i}^a(\t_1) \psi_{i}^b(\t_2) \right)^{q/2}
      \left( \sum_j \psi_{j}^c(\t_3) \psi_{j}^d(\t_4) \right)^{q/2}
    \right\rangle \,.
\end{align}
If $(a,b) \ne (c,d)$ then for some replica (say $a$) the fermions appear in the correlator all with the same time, and so this correlator vanishes (at leading order) on the replica-symmetric saddle.
Therefore only terms where $F_{ab}$ and $F_{cd}$ have the same replicas survive.
Let us set $a=c$, $b=d$, and consider a specific choice of $a,b$ with $a \ne b$.
\begin{align}
    \frac{N^{2-q}}{2} \int \! dt_{1,2,3,4} \,
    F_{ab}(\t_1,\t_2) F_{ab}(\t_3,\t_4)
    \left\langle
      \left( \sum_{i,j} \psi_{i}^a(\t_1) \psi_{i}^b(\t_2) \psi_{j}^a(\t_3) \psi_{j}^b(\t_4) \right)^{q/2}
    \right\rangle \,. \label{Fterm}
\end{align}
Let us now compute the correlator appearing in (\ref{Fterm}) 
at leading order in large $N$.
\begin{align}
    &\left\langle
      \left( \sum_{i,j} \psi_{i}^a(\t_1) \psi_{i}^b(\t_2) \psi_{j}^a(\t_3) \psi_{j}^b(\t_4) \right)^{q/2}
    \right\rangle \nonumber \\
    = \, &
    (-1)^{q/2} 
    \sum_{\vec{i},\vec{j}}
    \left\langle
      \psi_{i_1}^a(\t_1) \psi_{j_1}^a(\t_3) \cdots
      \psi_{i_{q/2}}^a(\t_1) \psi_{j_{q/2}}^a(\t_3) \cdot
      \psi_{i_1}^b(\t_2) \psi_{j_1}^b(\t_4) \cdots
      \psi_{i_{q/2}}^b(\t_2) \psi_{j_{q/2}}^b(\t_4)
    \right\rangle \nonumber\\
    = \, &
    (-1)^{q/2} 
    \sum_{\vec{i},\vec{j}}
    \left\langle
      \psi_{i_1}(\t_1) \psi_{j_1}(\t_3) \cdots
      \psi_{i_{q/2}}(\t_1) \psi_{j_{q/2}}(\t_3)
    \left\rangle
      \cdot
    \right\langle
      \psi_{i_1}(\t_2) \psi_{j_1}(\t_4) \cdots
      \psi_{i_{q/2}}(\t_2) \psi_{j_{q/2}}(\t_4)
    \right\rangle \nonumber\\
    = \, &
    (-1)^{q/2} \left( q/2 \right)! \, N^{q/2} \Big\langle G^{q/2}(\t_1,\t_3) \Big\rangle
    \Big\langle G^{q/2}(\t_2,\t_4) \Big\rangle + \cdots
    \,.
\end{align}
In the last step we kept only the diagonal terms, which give the leading contribution at large $N$.
Plugging this in \eqref{Fterm} and then \eqref{Fquad}, we find
\begin{align}
  - \int m_{ab}^2 F_{ab}^2 &=
    - \frac{qN}{J^2} \int d\t_1 d\t_2 F_{ab}^2(\t_1,\t_2)
    \cr &\quad
    + (-1)^{q/2} \left( q/2 \right)! \, N^{2-q/2} \int \! dt_{1,2,3,4} \,
    F_{ab}(\t_1,\t_2) F_{ab}(\t_3,\t_4)
    \Big\langle G^{q/2}(\t_1,\t_3) \Big\rangle
    \Big\langle G^{q/2}(\t_2,\t_4) \Big\rangle
    \,. \cr 
\end{align}
Focusing on the time-independent mode of $F_{ab}$, we get equation \eqref{mab} as advertised.

\section{Numerical methods}
\label{app:numerics}

In this appendix we provide details about the numerical methods used to compute the results of Section~\ref{numerics}.
We used two independent implementations to test our results, one running on CPUs and one on GPUs.
The GPU implementation can be found at \hyperlink{https://github.com/guygurari/syk}{https://github.com/guygurari/syk}.

For a system consisting of $2N$ Majorana fermions, the Hilbert space is $2^N$-dimensional (in the rest of the paper we denote the number of Majorana fermions by $N$). Implementing the $Z_2$ symmetry associated with the Majorana fermion parity conservation, it reduces down to $2^{N-1}$. For large values of $N$, constructing the Hamiltonian operator and working with it becomes exponentially harder, and as a result the exact diagonalization of such systems becomes unfeasible beyond some $N$. However, it is possible to employ a simple trick widely used in Density-Matrix-Renormalization-Group (DMRG) and related approaches to increase the largest accessible values of $N$. This trick reduces RAM consumption from $2^{N}$ to $c_{N}2^{N/2}$, where $c_N$ grows polynomially with $N$ ($N^2$ in the case of SYK model) and thus the space complexity of the diagonalization algorithm is reduced significantly. 

The main observation in this method is that the system can be divided into left and right subsystems, with $2N_L$ and $2N_R$ Majorana fermions, respectively where $N_R = N - N_L$. The Hilbert space associated with the $L$ ($R$) subsystem is now $D_L = 2^{N_L}$ ($D_R = 2^{N_R}$) dimensional. On the other hand, the total Hamiltonian can in general be written as the following Schmidt decomposition:
\begin{align}
\label{eq:SD-Hamiltonian}
H = H_{L}\otimes \mathbb{I}_R + \mathbb{I}_L \otimes H_{R} + \sum_{a}g_a \mathcal{O}_{L}^{a} \otimes \mathcal{O}_{R}^{a} \,,
\end{align}
with properly chosen $\mathcal{O}_{L/R}^{a}$ operators and $g_a$ couplings. Computing the tensor product operations in the above representation will bring us back to the standard approach to exact diagnoalization. However, tensor product operations are quite expensive computationally and storing the resulting huge matrices is costly. It is possible to avoid doing such unnecessary costly operations and still perform diagonalization algorithms efficiently. 

The Lanczos algorithm is one of the most popular methods for obtaining low energy eigenvectors and eigenvalues of sparse matrices \cite{cullum2002lanczos}.
It is a power iteration method based on successive matrix-vector multiplication operations, $v_{b+1} =  H v_{b}$, starting from an initial random vector $v_0$. The resulting $v_b$'s form basis vectors for the Lanczos diagonalization procedure. The desired vector operations can be implemented more efficiently using the Schmidt decomposition of $v_b$ vectors, namely:
\begin{align}
\label{eq:SD-vector}
v_{b} = \sum_{a} \lambda^{(b)}_a v^{(b)}_{a,L} \otimes v^{(b)}_{a,R} \,.
\end{align}
where $v^{(b)}_{a,L/R}$ are orthogonal basis vectors defined on the $L$ ($R$) subsystem. We then utilize the following unitary (duality) transformation on the right side : $v^{(b)}_{a,R} \to v^{(b)~{\rm T}}_{a,R}$ which in turn yields
\begin{align}
\label{eq:reshaped-vec}
v_{b} \to \overline{v}_{b} = {\rm reshape}\left(v_{b},D_L,D_R\right) \,.
\end{align}
transformation on the $v_{b}$ vectors. The transformed $v_{b}$, $\overline{v_b}$, is now a $D_L \times D_R$ dimensional matrix. Next, we consider $v_{b+1} = H v_{b}$. It can be verified that $\overline{v}_{b+1}= {\rm reshape}\left(v_{b},D_L,D_R\right)$ can be evaluated using the relation
\begin{align}
\label{eq:H-v-operation}
\overline{v}_{b+1} =  H_L\overline{v}_{b} + \overline{v}_{b}H_R^{\rm T} + \sum_{a} g_{a}\mathcal{O}_L^{a} \overline{v}_{b}\mathcal{O}_R^{a~\rm T}\, .
\end{align}
This way we never need to explicitly compute the tensor products $\mathcal{O}^a_L\otimes \mathcal{O}^a_R$, and instead we just need to store the $\mathcal{O}^a_{L/R}$ operators on the RAM. For $N_L = N_R = N/2$, this approach requires storing $2^{N/2}$-dimensional matrices on the RAM, and there are $O(N^2)$ such operators that take part in the interaction between the left and right subsystems of the SYK model. Hence, the space complexity of this approach is $O(N^2 2^{N/2})$ for the SYK model instead of $O(2^N)$ of the conventional Lanczos method. It is worth mentioning that the space complexity affects the computation time and the above procedure can reduce it by orders of magnitude. Furthermore, the above trick can change the time complexity of the Lanczos algorithm (and in particular of the main step $v \to Hv$) from $O(D_L^2 D_R^2)$ down to $O(D_LD_R(D_L+D_R))$ when $O_L^{a}$ and $O_R^{a}$ operators are dense matrices.
For sparse operators, the time complexity is unaffected by the above scheme. However, we have noticed that in practice the complexity can drop significantly using the above method, especially for long range Hamiltnonians such as SYK (indeed, the overall prefactor of the time complexity decreases).

We implemented the Lanczos algorithm as described on GPUs, taking advantage of their ability to carry out highly parallel calculations.
In common implementations of the Lanczos algorithm one keeps track of previously computed eigenvectors, in order to overcome the inherent numerical instabilities of the algorithm.
Despite the lower space complexity described above, this method is still too costly to run on GPUs due to their relatively limited RAM.
Instead, in the GPU code we used an alternative implementation of the Lanczos algorithm which does not need to keep track of the eigenvectors \cite{cullum2002lanczos}.
This method is useful when one is only interested in the eigenvalues of the matrix.

\section{The Quantum Sherrington-Kirkpatrick model}
\label{app:sk}

In this appendix, we present numerical calculations of the eigenvalue statistics in the quantum Sherrington-Kirkpatrick model \cite{PhysRevLett.70.3147}.
The Hamiltonian is that of the transverse field Ising model on $N$ sites, but with random infinite-range couplings:
\begin{align}
\label{eq:HquantumSK}
H = \sum_{ij} J_{ij} X_i X_j + \Gamma\sum_i Z_i \,,
\end{align}
where $X_i$ and $Z_i$ are the Pauli-$X$ and Pauli-$Z$ matrices on site $i$.
The sum in the first term runs over all pairs in the system, 
and the couplings $J_{ij}$ are independent Gaussian random variables with mean zero and variance $\langle J_{ij}^2 \rangle = 1/N$.
This Hamiltonian has a $\mathbb{Z}_2$ symmetry represented by the unitary operator $U = Z_1 \ldots Z_N$.

\begin{figure}[!h]
    \centering
    \includegraphics[width=0.3\textwidth]{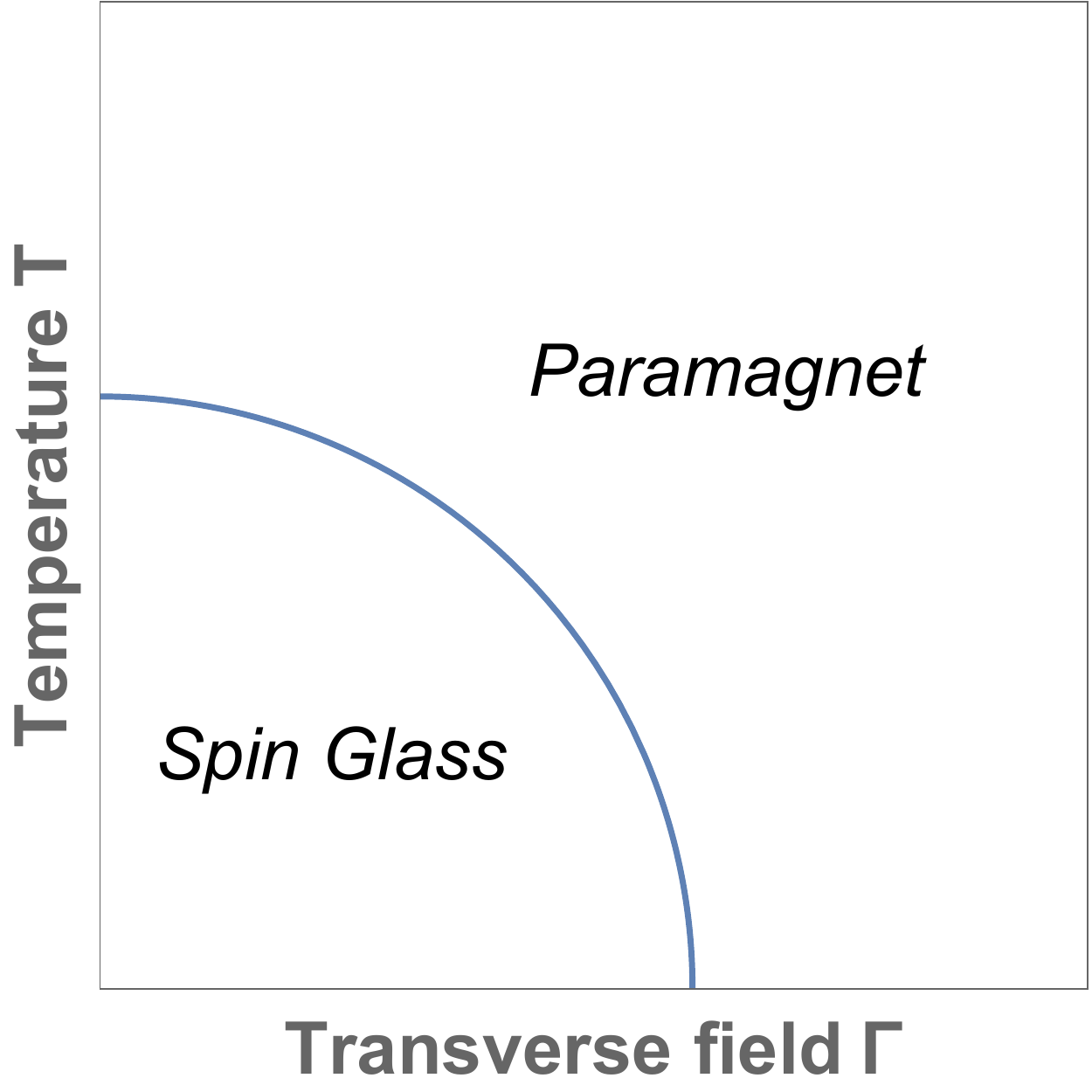}
    \caption{Schematic phase diagram of the quantum Sherrington-Kirkpatrick model. There is a spin glass phase at small external field $\Gamma$ and small temperatures.}
    \label{fig:quantumSKphasediagram}
\end{figure}

When $\Gamma=0$ in the Hamiltonian \eqref{eq:HquantumSK}, all the terms in the Hamiltonian commute and the model reduces to the classical Sherrington-Kirkpatrick model, which is well-known to have a spin glass phase \cite{sherrington, parisi, mezard1987spin} at low temperatures.
The spin glass phase persists at small $\Gamma$.
The spin glass phase can be destroyed by either increasing $\Gamma$ or by increasing the temperature beyond their critical values.
A cartoon phase diagram of the model is shown in Figure \ref{fig:quantumSKphasediagram}.

\begin{figure}[h!]
    \centering
    \includegraphics[scale=0.47]{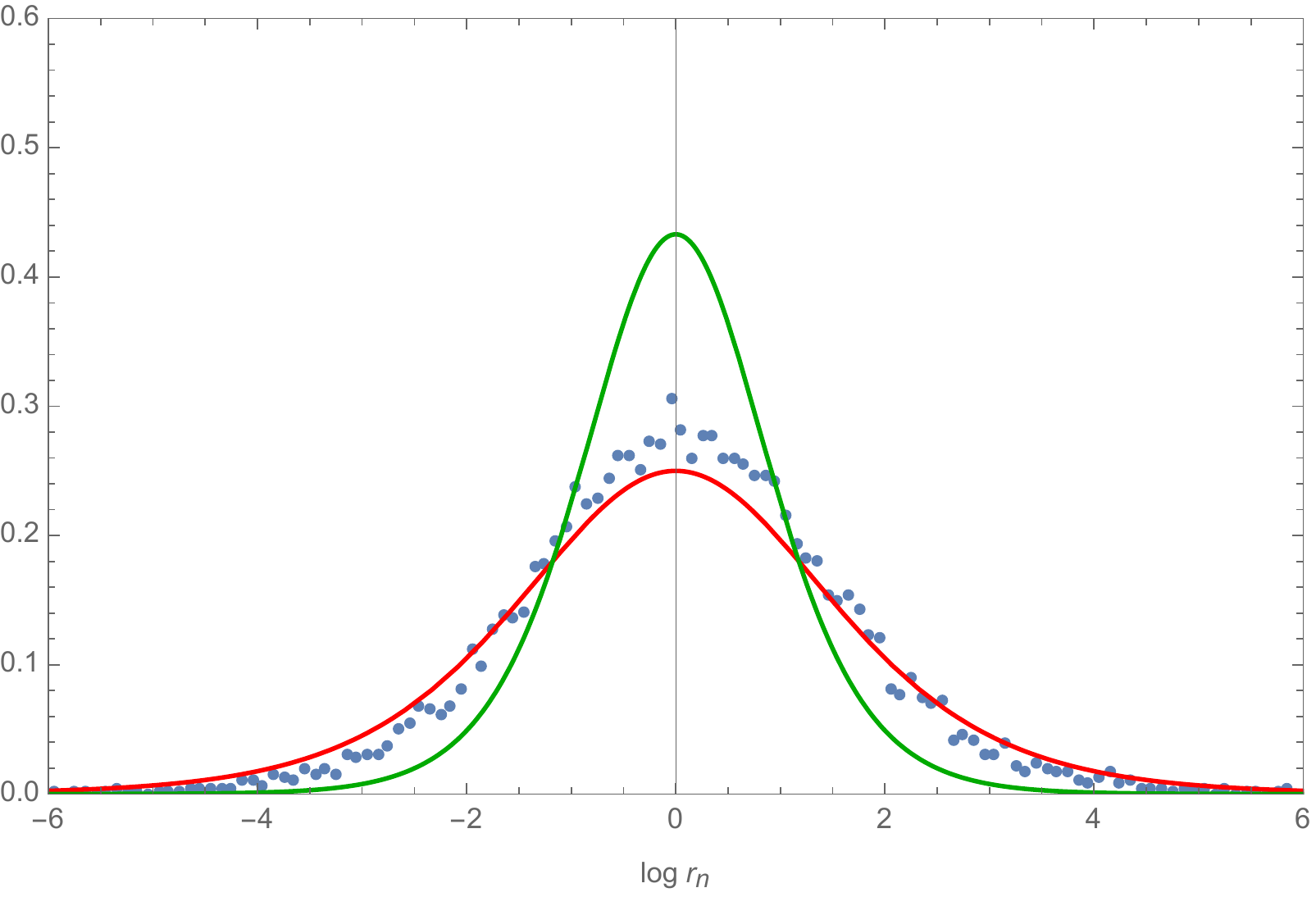}
    \includegraphics[scale=0.47]{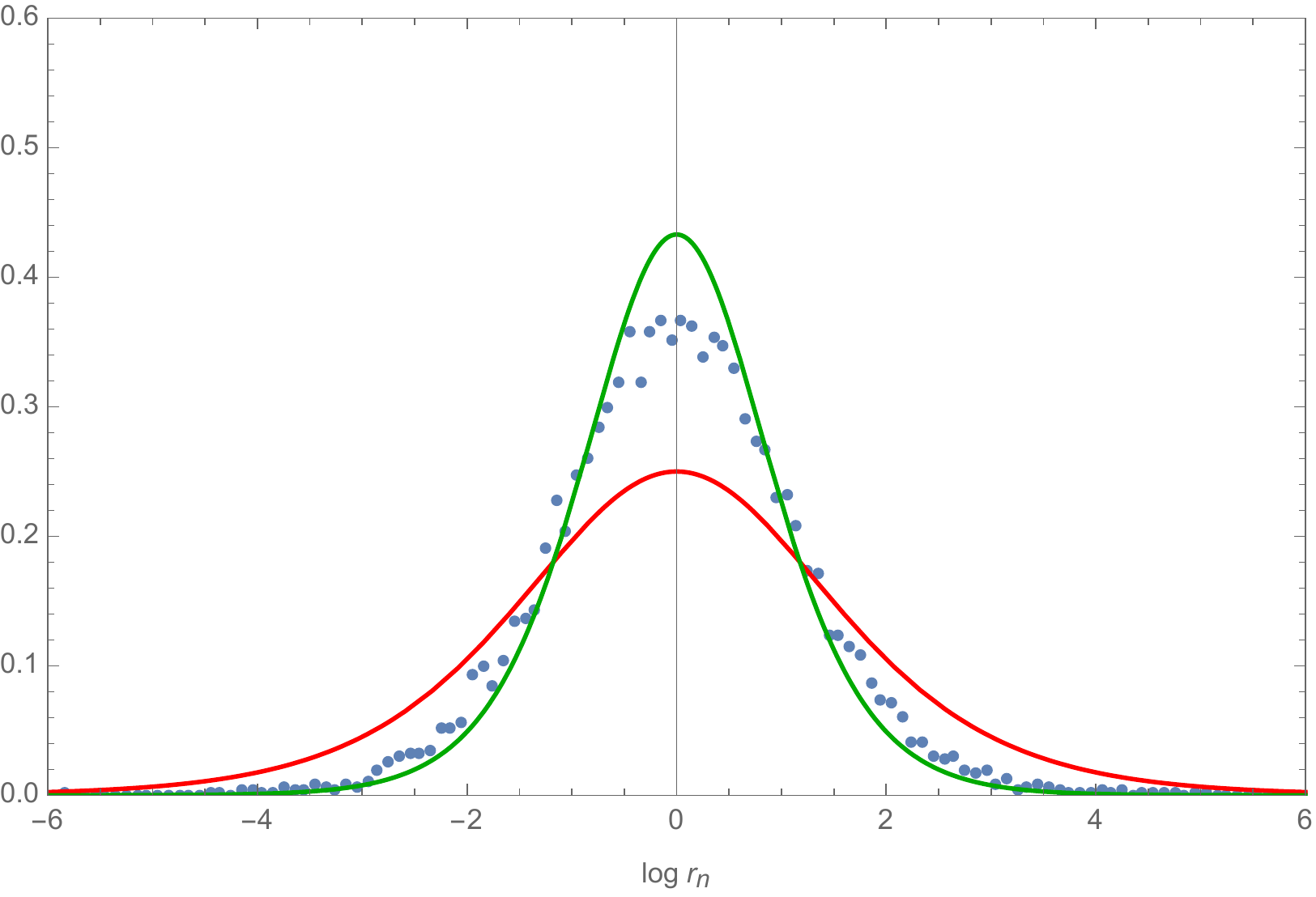}
    \caption{Distribution of $\log r_n$ when $\Gamma = 0.1$ for low lying states (left) and for states in the middle of the spectrum (right). The blue dots are numerical data, the red curve is the exponential distribution, and the green curve is the GOE ensemble prediction. There is a spin glass phase at low temperatures, and consequently the distribution is exponential. Since the spin glass phase is wiped out at high temperatures, the states from the middle of the spectrum follow GOE statistics.}
    \label{fig:skspinglass}
\end{figure}

We project the Hamiltonian to the even $\mathbb{Z}_2$ sector and perform exact diagonalization on a system of $N=12$ spins.
The level spacing statistics calculation was described in Section~\ref{sec:lss} (see also \cite{You:2016ldz}).
In Figure \ref{fig:skspinglass}, we show the distribution of $\log r_n$ when $\Gamma = 0.1$.
At this value of $\Gamma$, the low temperature phase is a spin glass.
Thus the low energy part of the spectrum should exhibit exponential statistics, as is clearly visible in the left panel of Figure \ref{fig:skspinglass}.
We use 200 disorder realizations and the lowest 50 states from each realization.
The high temperature phase is ergodic, and thus states drawn from the middle of the spectrum should exhibit GOE statistics.
This is also clearly visible in the right panel of Figure \ref{fig:skspinglass}.
Here we take 200 disorder realizations and the middle 50 states from the spectrum of each realization. 

\begin{figure}[h!]
    \centering
    \includegraphics[width=0.5\textwidth]{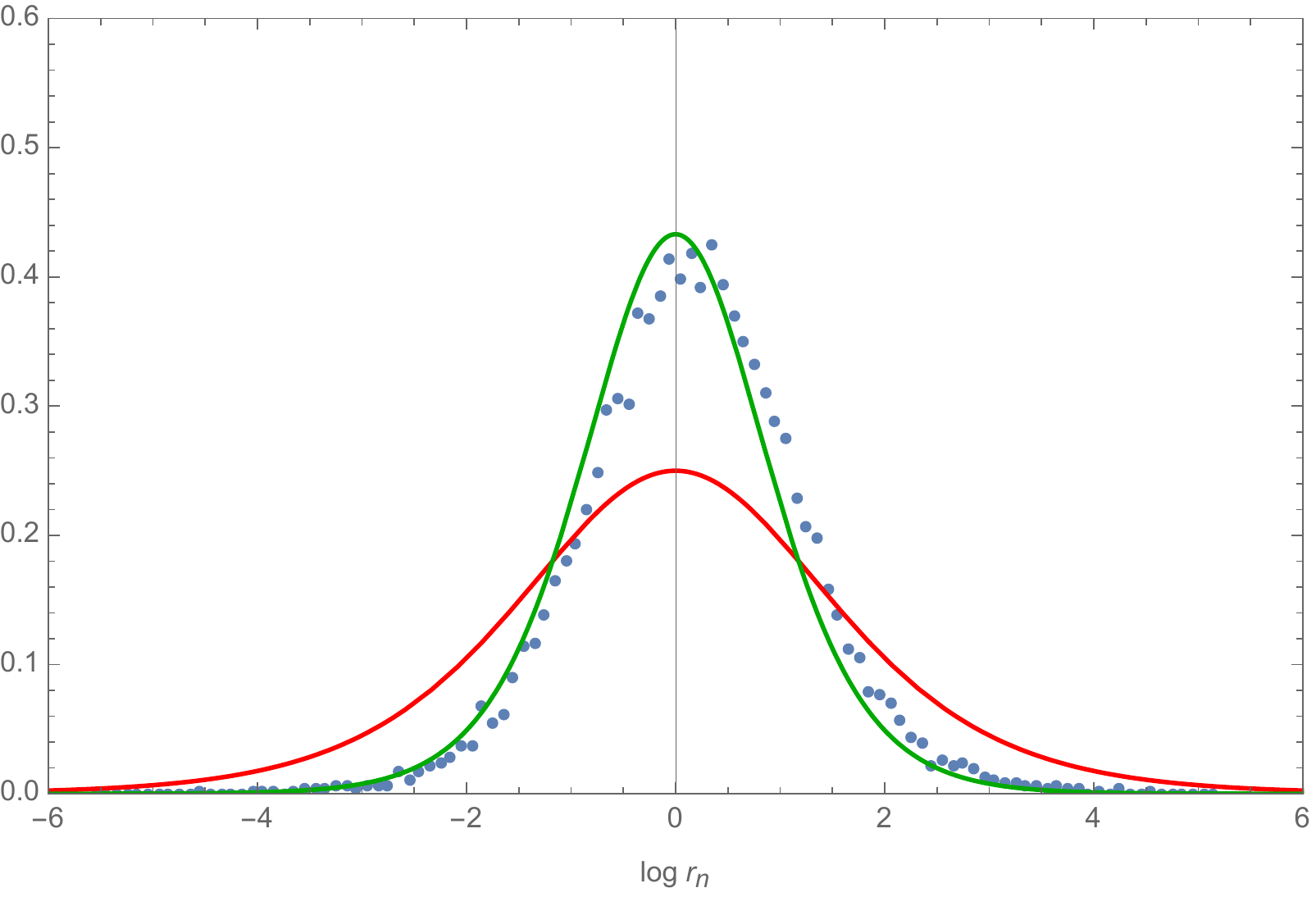}
    \caption{Distribution of $\log r_n$ for the low lying states when $\Gamma = 5.0$. There is no spin glass phase and thus even the low-lying states follow GOE statistics. The blue dots are numerical data, the red curve is for the exponential distribution, and the dark green curve is for the GOE ensemble.}
    \label{fig:skparamagnetic}
\end{figure}

Finally, consider the case where $\Gamma$ is large. 
Here there is no spin glass phase at any temperature, so even the low energy part of spectrum should exhibit GOE statistics.
This is confirmed in Figure~\ref{fig:skparamagnetic}.

\newpage
\bibliography{syk-at-low-energies.bib}
\bibliographystyle{JHEP-2}

\end{document}